\documentclass[12pt]{iopart}
\usepackage{iopams}
\usepackage{graphicx}

\begin{document}
\title[TAP equation for analog neural network with temporally fluctuating synaptic noise]
{Thouless-Anderson-Palmer equation for analog neural network with temporally fluctuating white synaptic noise}
\author{Akihisa Ichiki and Masatoshi Shiino}
\address{Department of Applied Physics, Faculty of Science, Tokyo Institute of Technology, 2-12-1 Ohokayama Meguro-ku Tokyo, Japan}
\ead{aichiki@mikan.ap.titech.ac.jp}
\begin{abstract}
Effects of synaptic noise on the retrieval process of associative memory neural networks are studied from the viewpoint of neurobiological and biophysical understanding of information processing in the brain.
We investigate the statistical mechanical properties of stochastic analog neural networks with temporally fluctuating synaptic noise, which is assumed to be white noise. 
Such networks, in general, defy the use of the replica method, since they have no energy concept. 
The self-consistent signal-to-noise analysis (SCSNA), which is an alternative to the replica method for deriving a set of order parameter equations, requires no energy concept and thus becomes available in studying networks without energy functions. 
Applying the SCSNA to stochastic network requires the knowledge of the Thouless-Anderson-Palmer (TAP) equation which defines the deterministic networks equivalent to the original stochastic ones. 
The study of the TAP equation which is of particular interest for the case without energy concept is very few, while it is closely related to the SCSNA in the case with energy concept. 
This paper aims to derive the TAP equation for networks with synaptic noise together with a set of order parameter equations by a hybrid use of the cavity method and the SCSNA. 
\end{abstract}


\submitto{\JPA}
\maketitle

\section{Introduction}
The replica method \cite{SK} for random spin systems has been successfully employed in neural network models of associative memory \cite{MPV,AGS}. 
However the replica method requires the concept of free energy. 
On the other hand, various types of neural network models which have no energy concept, such as a network with asymmetric synaptic coupling or temporally fluctuating synaptic coupling, may be in existence. 
The self-consistent signal-to-noise analysis (SCSNA) \cite{Shiino, Shiino2, Shiino3}, which is an alternative approach to the replica method for deriving a set of order parameter equations, requires no energy concept. 
Thus it can be applicable to study the statistical properties of wider class of networks including networks without energy concept. 

The SCSNA, which was originally proposed for deriving a set of order parameter equations for a deterministic analog neural network with or without energy functions, becomes applicable to a stochastic network by noting that the TAP equation defines the deterministic one equivalent to the original stochastic one. 
The SCSNA is closely related to the Thouless-Anderson-Palmer (TAP) equation \cite{TAP, Morita} via the concept of cavity method in the case where a network has energy concept \cite{Shiino3} and the relationship between the two was studied in detail in the networks with two-body and multi-body interactions \cite{Shiino3, Ichiki}. 
The coefficient of the Onsager reaction term characteristic to the TAP equation which determines the form of the transfer function is self-consistently obtained through the concept shared by the cavity method and the SCSNA. 

The TAP equation is of our interest for studying the statistical properties of a network without energy concept. 
However the TAP equations for a network with synaptic noise are not found in literatures. 
The main target of this paper is to derive the TAP equation for a network with temporally fluctuating synaptic noise as {\it multiplicative noise}. 

The effects of such synaptic noise on the retrieval properties of networks have been studied in some recent works \cite{Garrido, Marro, Garrido2, Torres}. 
According to the stochastic resonance theory \cite{Gammaitoni, Benzi, Nicolis}, the temporally fluctuating synaptic noise may possibly be  expected to reduce the interference of the uncondensed patterns on the retrieval property of the network (noise in terms of stochastic resonance) and, as a result, enhance the retrieval property or the storage capacity (signal) of the network. 
However such an argument is not found in literatures. 
To discuss the effect of temporally fluctuating synaptic noise on the retrieval property, it is important to construct a tractable model to study the role played by such synaptic noise. 

In literatures, the term "synaptic noise" is used in three meanings: (i) quenched disorder in synaptic couplings \cite{Sompolinsky}, (ii) randomness related to the dilution of synaptic couplings \cite{Treves}, and (iii) temporal fluctuation of synaptic couplings \cite{Garrido, Marro, Garrido2, Torres}. 
We will use the term "synaptic noise" in the third meaning in the present paper. 

Cortes \etal \cite{Cortes} (see also \cite{Torres}) investigated the case where neurons obey a master equation with continuous time and the synaptic couplings obey (i) slow dynamics \cite{Amari, Hopfield}, (ii) fast dynamics \cite{Garrido2, Torres, Gardiner, Cortes} (iii) middle speed dynamics \cite{Pantic, Cortes2} compared to the dynamics of neurons. 
In the first case, since the synaptic couplings obey slow dynamics, the adiabatic approximation for the synaptic couplings becomes exact in the limit where the time scale of synapse dynamics $\tau\to\infty$ and quenched random noise in couplings again arises. 
Thus the synaptic noise is regarded as the well-known quenched random
variable and this type of synaptic noise has been studied as "synaptic noise" in many literatures \cite{Choi}. 

In the second case, since the dynamics of the synaptic noise is sufficiently fast compared to the dynamics of neurons, one can define the effective strength of synaptic coupling by averaging the temporally fluctuating synaptic coupling \cite{Garrido, Marro, Garrido2, Torres, Marro2}. 
In this case, one can find the effective Hamiltonian of the network and use well-known replica method \cite{SK} to obtain the order parameters analytically \cite{MPV, AGS}. 
Other example of the fast synaptic dynamics can be found in \cite{Uezu}, which studies the properties of the equilibrium state of the system with stochastically evolving couplings. 

On the other hand, the third case is difficult to deal with analytically especially in the case where the number of memory patterns is proportional to the total number of neurons and only numerical results based on computer simulations exist \cite{Pantic, Cortes2}. 
In spite of these recent efforts to elucidate the effects of synaptic noise on the retrieval properties of neural networks, such preceding studies have been based on the macroscopic viewpoint, where the order parameters solely have been investigated, and the TAP equations for such cases have not been reported. 

The purpose of this paper is two-fold: (i) we will derive the TAP equation for a stochastic analog network with temporally fluctuating multiplicative synaptic noise which is not found in literatures. 
(ii) We will study the SCSNA and the TAP equation for such a network to elucidate the effects of the multiplicative synaptic noise on the retrieval property from both microscopic and macroscopic viewpoint. 
Part of this work is reported elsewhere \cite{Ichiki2}. 

This paper is organized as follows: in the next section, we will describe an analog neural network model with temporally fluctuating synaptic noise as multiplicative noise which is assumed to be white noise to write down a set of Langevin equations, 
and derive the corresponding Fokker-Planck equation. 
We will see that the equilibrium solution of the Fokker-Planck equation is given as a Gibbs probability density with the effective temperature, which should be determined self-consistently in the thermodynamic limit. In section 3, we will apply the cavity method to derive the formal expression of the TAP equation (pre-TAP equation). 
Then using the SCSNA, we will self-consistently obtain the concrete form of the transfer function which yields the complete form of the TAP equation as well as a set of order parameter equations. In section 4, the phase diagram for our model will be shown. In the last section, we will conclude this paper. 

\section{Model and Fokker-Planck equation formalism}
Let us deal with the following stochastic analog neural network of $N$ neurons with temporally fluctuating synaptic noise: 
\numparts
\begin{eqnarray}
\dot{x}_{i} = -\phi^{\prime}(x_{i}) &+& \displaystyle\sum_{j(\neq i)}
J_{ij}(t)x_{j}+\eta_{i}(t)\, ,\label{1}\\
\langle\eta_{i}(t)\eta_{j}(t^{\prime})\rangle &=& 2D\delta_{ij}\delta 
(t-t^{\prime})\, ,
\end{eqnarray}
\endnumparts
where $x_{i}$ ($i=1,\cdots ,N$) represents a state of the neuron at site $i$ taking a continuous value, $\phi(x_{i})$ is a potential of an arbitrary form which determines the probability distribution of $x_{i}$ in the case without the input $\sum_{j(\neq i)} J_{ij}x_{j}$, $\eta_{i}$ the Langevin white noise with its noise intensity $2D$ and $J_{ij}(t)$ the synaptic coupling. 
We note here that, in the case of associative memory neural network, the synaptic coupling $J_{ij}$ is usually defined by the well-known Hebb learning rule. 
However some experimental results show that the synaptic couplings have temporal fluctuations which originate from the dynamics of neurotransmitters or kinetics of ion channels independent of that of neurons \cite{Anderson}, and hence the effects of such synaptic noise may be relevant to the retrieval properties in realistic networks. 
To investigate such effects of synaptic noise, we assume the synaptic coupling taking the form: 
\numparts
\begin{eqnarray}
& &J_{ij}(t) = \bar{J}_{ij} + \epsilon_{ij}(t)\, ,\\
& &\langle\epsilon_{ij}(t)\epsilon_{kl}(t^{\prime})\rangle = 
\frac{2\tilde{D}}{N}\delta_{ik}\delta_{jl}\delta (t-t^{\prime})\, ,
\end{eqnarray}
\endnumparts
where $\bar{J}_{ij}$ is defined by the usual Hebb learning rule $\bar{J}_{ij}\equiv\frac{1}{N}\sum_{\mu =1}^{p}\xi_{i}^{\mu}\xi_{j}^{\mu}$ with $p=\alpha N$ the number of patterns embedded in the network, $\xi_{i}^{\mu}=\pm 1$ is the $\mu^{\mathrm{th}}$ embedded pattern at neuron $i$, and $\epsilon_{ij}(t)$ denotes the synaptic noise independent of $\eta_{i}(t)$, which we assume in our model as white noise with its intensity $2\tilde{D}/N$ for simplicity. 
Notice that, in equation (\ref{1}), the synaptic noise behaves as {\it multiplicative noise} and the synaptic coupling $J_{ij}(t)$ is asymmetric. 

Noting 
\begin{eqnarray}
\fl\displaystyle\lim_{\Delta t\to 0}\frac{1}{\Delta t}
\int_{t}^{t+\Delta t}\, ds\int_{t}^{t+\Delta t}\, ds^{\prime}
\left\langle\sum_{k(\neq i)}\epsilon_{ik}(s)x_{k}(s)\sum_{l(\neq j)}
\epsilon_{jl}(s^{\prime})x_{l}(s^{\prime})\right\rangle = 
\frac{2\tilde{D}}{N}\delta_{ij}\sum_{k(\neq i)}x^{2}_{k}\, \nonumber
\end{eqnarray}
by means of Ito integral, we obtain the Fokker-Planck equation corresponding to the Langevin equation (\ref{1}) as 
\begin{eqnarray}
\fl\frac{\partial P(t,\mathbf{x})}{\partial t} = -\displaystyle
\sum_{i=1}^{N}\frac{\partial}{\partial x_{i}}\left\{ 
-\phi^{\prime}(x_{i}) + \sum_{j(\neq i)}\bar{J}_{ij}x_{j}
- \left( D+\tilde{D}\hat{q}\right)
\frac{\partial}{\partial x_{i}}\right\}P(t,\mathbf{x})\, ,\label{FPE}
\end{eqnarray}
where $\hat{q}\equiv\frac{1}{N}\sum_{j(\neq i)}x_{j}^{2}$. 
Since the self-averaging property holds in the thermodynamic limit $N\to\infty$, one can identify $\hat{q}$ as 
\begin{eqnarray}
\hat{q}=\frac{1}{N}\displaystyle\sum_{i=1}^{N}
\langle x_{i}^{2}\rangle\, ,\label{QHAT}
\end{eqnarray}
where $\langle\cdot\rangle$ represents the thermal average with respect to $P(t,\mathbf{x})$. 
Thus equation (\ref{FPE}) is found to be a nonlinear Fokker-Planck equation whose diffusion coefficient $D+\tilde{D}\hat{q}$ depends on the probability density $P(t,\mathbf{x})$ 
\cite{Frank, Shiino4}. 
In this paper we are concerned with deriving the TAP equation and order parameter equations for the equilibrium state self-consistently. 
Furthermore the order parameter $\hat{q}$ is also obtained self-consistently in our framework as seen below. 
Supposing $\hat{q}$ is given, the Fokker-Planck equation (\ref{FPE}) turns to be a linear equation and one can easily find the equilibrium probability density for the linear Fokker-Planck equation (\ref{FPE}) as 
\begin{eqnarray}
P_{N}({\mathbf x})=Z^{-1}\exp\left\{ -\beta_{\mathrm{eff}}\left(
\displaystyle\sum_{i=1}^{N}\phi (x_{i})-\sum_{i<j}
\bar{J}_{ij}x_{i}x_{j}\right)\right\}\, ,\label{eqprob}
\end{eqnarray}
where $Z$ denotes the normalization constant and 
\begin{eqnarray}
\beta_{\mathrm{eff}}^{-1}\equiv D+\tilde{D}\hat{q} \label{5.5}
\end{eqnarray} 
plays the role of the {\it effective temperature of the network}. 
Notice that the temperature of the system is modified to $\beta_{\mathrm{eff}}^{-1}$ as a consequence of the multiplicative noise and it depends on the order parameter $\hat{q}$. 
Here it is easily checked that the equilibrium distribution of the system becomes Gibbs distribution in the thermodynamic limit $N\to\infty$. 

Since we have explicitly written down the equilibrium probability distribution density (\ref{eqprob}) as a form of Gibbs distribution, one can define the (effective) Hamiltonian of $N$-body system as 
\begin{eqnarray}
H_{N}\equiv \displaystyle\sum_{i=1}^{N}\phi (x_{i}) - 
\sum_{i<j}\bar{J}_{ij}x_{i}x_{j}\, .\label{hamiltonian}
\end{eqnarray}
Then regarding the original network with multiplicative noise (\ref{1}) as an analog version of the standard Hopfield model whose Hamiltonian is given by equation (\ref{hamiltonian}) with the effective temperature $\beta_{\mathrm{eff}}^{-1}$, one can apply the usual cavity method \cite{MPV} to this system and derive the (pre-)TAP equation. 

\section{Cavity method and self-consistent signal-to-noise analysis}
We have obtained the equilibrium probability density as a form of Gibbs distribution (\ref{eqprob}) and the effective Hamiltonian (\ref{hamiltonian}) in the previous section. 
Thus the cavity method \cite{MPV}, which is usually applied to the network models for deriving the TAP equation, is applicable for our model. 
According to the cavity method, we divide the Hamiltonian of $N$-body system (\ref{hamiltonian}) into that of $(N-1)$-body system and the part involving in the state of $i^{\mathrm{th}}$ neuron as 
\begin{eqnarray}
H_{N}=\phi (x_{i})-h_{i}x_{i} + H_{N-1}\, ,\nonumber
\end{eqnarray}
where $h_{i}\equiv\sum_{j(\neq i)}\bar{J}_{ij}x_{j}$ is the local field at site $i$ and the Hamiltonian of $(N-1)$-body system $H_{N-1}$ is given as 
$H_{N-1}\equiv \sum_{j(\neq i)}\phi (x_{j}) - \sum_{j<k(\neq i)}\bar{J}_{jk}x_{j}x_{k}$. 
Then the marginal probability density distribution of $x_{i}$ and the local field $h_{i}$ is given as 
\begin{eqnarray}
P_{N}(x_{i},h_{i}) &=& \int\left[\displaystyle
\prod_{j(\neq i)}dx_{j}
\right]\delta\left( h_{i}-\sum_{j(\neq i)}\bar{J}_{ij}x_{j}
\right) P_{N}({\mathbf x})\nonumber\\
&=&\tilde{Z}^{-1}\exp\left\{\ -\beta_{\mathrm{eff}}\left[
\phi (x_{i})-h_{i}x_{i}\right]\right\} P_{N-1}(h_{i})\, ,\nonumber
\end{eqnarray}
where $\tilde{Z}$ is the normalization constant and $P_{N-1}(h_{i})$ denotes the probability density of the local field $h_{i}$ in the $(N-1)$-body system defined as 
\begin{eqnarray}
P_{N-1}(h_{i})\equiv Z_{N-1}^{-1}\int\left[\displaystyle\prod_{j(\neq i)}
dx_{j}\right]\delta\left( h_{i}-\sum_{j(\neq i)}\bar{J}_{ij}
x_{j}\right)\exp\left[ -\beta_{\mathrm{eff}}H_{N-1}\right]\, ,\nonumber
\end{eqnarray}
where $Z_{N-1}$ denotes the normalization constant. Since the local field is given as the summation of a sufficiently large number of random variables and their cross-correlations are expected to be $\mathrm{O}(1/\sqrt{N})$, one can expect that $P_{N-1}(h_{i})$ turns out to be a Gaussian density in the thermodynamic limit $N\to\infty$ according to the central limit theorem: 
\begin{eqnarray}
P_{N-1}(h_{i})=\frac{1}{\sqrt{2\pi\sigma^{2}}}\exp\left[ -
\frac{(h_{i}-\langle h_{i}\rangle_{N-1})^{2}}{2\sigma^{2}}
\right]\, ,\nonumber
\end{eqnarray}
where $\langle\cdot\rangle_{N-1}$ represents the thermal average with respect to the $(N-1)$-body probability density $P_{N-1}(\mathbf{x})$ and $\sigma^{2}$ is the variance of $P_{N-1}(h_{i})$, which is evaluated later self-consistently in the framework of the SCSNA. 
Then taking the average of $x_{i}$ with respect to the marginal probability $P_{N}(x_{i},h_{i})$ straightforwardly yields 
\begin{eqnarray}
\langle x_{i}\rangle = F(\langle h_{i}\rangle_{N-1})\, ,
\label{Nm1}
\end{eqnarray}
where $F$ is a transfer function defined as 
\begin{eqnarray}
F(y)\equiv\frac{\int dx\,x\exp\left\{ -\beta_{\mathrm{eff}}
\left[\phi (x)-yx-\frac{\beta_{\mathrm{eff}}\sigma^{2}}{2}
x^{2}\right]\right\}}{\int dx\,\exp\left\{ 
-\beta_{\mathrm{eff}}
\left[\phi (x)-yx-\frac{\beta_{\mathrm{eff}}\sigma^{2}}{2}x^{2}
\right]\right\}}\, .\label{F}
\end{eqnarray}
Similarly $\langle h_{i}\rangle_{N-1}$ is obtained as 
\begin{eqnarray}
\langle h_{i}\rangle_{N-1}=\langle h_{i}\rangle -
\beta_{\mathrm{eff}}\sigma^{2}\langle x_{i}\rangle\, .\nonumber
\end{eqnarray}
Thus we have the pre-TAP equation 
\begin{eqnarray}
\langle x_{i}\rangle = F\left( \displaystyle\sum_{j(\neq i)}
\bar{J}_{ij}\langle x_{j}\rangle -\Gamma_{\mathrm{Ons}}
\langle x_{i}\rangle\right)\, ,\label{preTAP}
\end{eqnarray}
where $\Gamma_{\mathrm{Ons}}\equiv\beta_{\mathrm{eff}}\sigma^{2}$. 
Since the concrete form of the transfer function $F$ depends on the effective temperature $\beta_{\mathrm{eff}}$ and the variance of the local field $\sigma^{2}$, it is necessary to obtain $\beta_{\mathrm{eff}}$ and $\sigma^{2}$ to have the TAP equation \cite{Shiino2, Shiino3}. 

Equation (\ref{preTAP}) is regarded as defining a deterministic analog network corresponding to the original stochastic one (\ref{1}), and hence we can apply the SCSNA to equation (\ref{preTAP}) to determine $\beta_{\mathrm{eff}}$ and $\sigma^{2}$ self-consistently as was studied for the case without synaptic noise \cite{Shiino2, Shiino3}. 
For simplicity, we here assume that the only one condensed pattern $\{\xi_{i}^{1}\}$ is retrieved. 
The extension to the case of an arbitrary finite number of condensed patterns is straightforward. 
Using the overlap order parameter $m^{\mu}\equiv \frac{1}{N}\sum_{i=1}^{N}\xi_{i}^{\mu}\langle x_{i}\rangle$, the equilibrium average of the local field is rewritten as 
\begin{eqnarray}
\langle h_{i}\rangle = \xi^{1}_{i}m^{1}+\displaystyle\sum_{\mu\ge 2}
\xi_{i}^{\mu}m^{\mu}-\alpha \langle x_{i}\rangle\, .\label{locfield}
\end{eqnarray}
Using the SCSNA, the above local field can be rewritten as \cite{Shiino2, Shiino3} 
\begin{eqnarray}
\langle h_{i}\rangle = \xi^{1}_{i}m^{1}+\xi^{\mu}_{i}m^{\mu}+
z_{i\mu}+\Gamma_{\mathrm{SCSNA}}\langle x_{i}\rangle\, ,\label{loc}
\end{eqnarray}
where $\sum_{\nu\ge 2}\xi_{i}^{\nu}m^{\nu}=\xi^{\mu}_{i}m^{\mu}+z_{i\mu}+\gamma\langle x_{i}\rangle$, $\Gamma_{\mathrm{SCSNA}}\equiv\gamma -\alpha$ and $z_{i\mu}$ is a Gaussian random variable with zero mean. 
As seen below, we will evaluate the overlap $m^{\mu}$ self-consistently, and then obtain $z_{i\mu}$ and $\gamma$ through the equivalence between the expression of the local field (\ref{locfield}) and (\ref{loc}). 
Substituting equation (\ref{loc}) into the pre-TAP equation (\ref{preTAP}) reads 
\begin{eqnarray}
\langle x_{i}\rangle =F\left(\xi_{i}^{1}m^{1}+\xi_{i}^{\mu}m^{\mu}
+z_{i\mu}+(\Gamma_{\mathrm{SCSNA}}-\Gamma_{\mathrm{Ons}})\langle x_{i}
\rangle\right)\nonumber
\end{eqnarray}
and comparing this equation with equation (\ref{Nm1}) yields \cite{Shiino3} 
\begin{eqnarray}
\Gamma_{\mathrm{SCSNA}}=\Gamma_{\mathrm{Ons}}\, ,\nonumber
\end{eqnarray}
since $\langle h_{i}\rangle_{N-1}$ is considered to be a Gaussian random variable which should not contain the Onsager reaction term. 
Noting that $m^{\mu}=\mathrm{O} (1/\sqrt{N})$ for $\mu\ge 2$, one can obtain the overlap for uncondensed patterns as 
\numparts
\begin{eqnarray}
m^{\mu}&=&\frac{1}{N(1-U)}\displaystyle\sum_{j=1}^{N}\xi_{j}^{\mu}
F(\xi^{1}_{j}m^{1}+z_{j\mu})\, ,\label{mmu}\\
U&\equiv& \frac{1}{N}\displaystyle\sum_{j=1}^{N}F^{\prime}(\xi^{1}_{j}
m^{1}+z_{j\mu})\, ,
\end{eqnarray}
\endnumparts
where $F^{\prime}$ denotes the derivative of the transfer function $F$ and the order expansion of $F$ with respect to $1/\sqrt{N}$ has been applied to $\langle x_{i}\rangle = F(\xi_{i}^{1}m^{1}+z_{i\mu}+\xi_{i}^{\mu}m^{\mu})$. 
Using equation (\ref{mmu}) and the definitions of $z_{i\mu}$ and $\gamma$, one finds 
\begin{eqnarray}
\gamma &=& \frac{\alpha}{1-U}\, ,\nonumber\\
z_{i\mu} &=& \frac{1}{N(1-U)}\displaystyle\sum_{\nu (\neq 1,\mu )}
\sum_{j(\neq i)}\xi_{i}^{\nu}\xi_{j}^{\nu}F(\xi^{1}_{j}m^{1}+z_{j\nu})
\, .\nonumber
\end{eqnarray}
Thus the variance of $z_{i\mu}$ is evaluated as 
\numparts
\begin{eqnarray}
\sigma_{z}^{2}=\frac{\alpha}{(1-U)^{2}}\left\langle F^{2}(
\xi m^{1}+z)\right\rangle_{\xi ,z}\, ,\label{sigma}
\end{eqnarray}
where $\langle\cdot\rangle_{\xi ,z}$ represents the average over a random variables $\xi =\pm 1$ and the Gaussian variable $z$, and the self-averaging property has been used. 
Similarly one obtains the set of order parameter equations as 
\begin{eqnarray}
m^{1}=\left\langle\xi F(\xi m^{1}+z)\right\rangle_{\xi ,z}\, ,\label{m}\\
U=\left\langle F^{\prime}(\xi m^{1}+z)\right\rangle_{\xi ,z}\, ,\label{U}\\
\Gamma_{\mathrm{Ons}}=\Gamma_{\mathrm{SCSNA}}=\frac{\alpha U}{1-U}\, .\label{gamma}
\end{eqnarray}
In the case where the multiplicative synaptic noise does not exist or the intensity of the synaptic noise is zero, i.e., $\beta_{\mathrm{eff}}=\beta\equiv 1/D$, the set of order parameter equations (\ref{sigma}), (\ref{m}), (\ref{U}), (\ref{gamma}) takes a closed form and determines the form of the transfer function $F$ as well as the order parameters self-consistently. 
For the case with multiplicative noise, however, it does not suffice to determine the form of the transfer function. 
We need the order parameter $\hat{q}$, which determines $\beta_{\mathrm{eff}}$, as well as $m^{1}$, $U$, $\sigma_{z}^{2}$, $\Gamma_{\mathrm{Ons}}$ to determine the concrete form of $F$. 
Since $\hat{q}$ is related to the macroscopic susceptibility of the system and, by definition of $F$ (\ref{F}), the order parameter $U$ corresponds with the susceptibility as $U=\beta_{\mathrm{eff}}(\langle x^{2}\rangle -\langle x\rangle^{2})$, one finds 
\begin{eqnarray}
\hat{q}=\frac{U}{\beta_{\mathrm{eff}}}+\frac{(1-U)^{2}}{\alpha}\sigma_{z}^{2}\, .\label{qhat}
\end{eqnarray}
\endnumparts
The set of equations (\ref{5.5}), (\ref{sigma}), (\ref{m}), (\ref{U}), (\ref{gamma}), (\ref{qhat}) takes a closed form and thus one can determine the form of $F$ self-consistently as well as the set of order parameters. 
Therefore substituting into the pre-TAP equation (\ref{preTAP}) the solutions $\beta_{\mathrm{eff}}$ and $\Gamma_{\mathrm{Ons}}$ that are self-consistently obtained within this framework 
yields the TAP equation. 

\section{Phase diagram and numerical results}
We have derived the TAP equation as well as the set of order parameter equations in the previous section. 
In this section we show the phase diagram by solving the set of order parameter equations (\ref{sigma}), (\ref{m}), (\ref{U}), (\ref{gamma}), (\ref{qhat}) numerically and investigate the effect of the multiplicative synaptic noise. 

For the well-known transfer function of the Ising neurons $F(x)=\tanh(\beta x)$, it is easy to understand the effects of the interference of the synaptic noise. 
This choice of the transfer function is equivalent to taking the potential $\phi$ as 
\begin{eqnarray}
\frac{\exp\left[ -\beta_{\mathrm{eff}}\phi (x)\right]}{\int\, \exp\left[
-\beta_{\mathrm{eff}}\phi (x)\right]\, dx} = 
\frac{1}{2}\delta (x-1)+\frac{1}{2}\delta (x+1)\, .\nonumber
\end{eqnarray}
In the Ising neuron model, since $\hat{q}$ is simply given as $\hat{q}=1$ and $\beta_{\mathrm{eff}}^{-1}=D+\tilde{D}$, the retrieval state vanishes for $\tilde{D}\ge 1$ according to the results of Amit-Geutfreund-Sompolinsky (AGS) \cite{AGS}. 

In this section, for simplicity, we consider the double-well potential whose minima are located at $x=\pm 1$: 
\begin{eqnarray}
\phi(x)=\frac{A}{4}x^{4}-\frac{A}{2}x^{2}\, ,\label{pot}
\end{eqnarray}
where $A$ determines the depth of the wells of the potential. 
This potential yields a continuous distribution of neuron states and thus defines an {\it analog} network model in which $\hat{q}$ is non-trivial. 
We investigate a phase diagram for the {\it analog} network model and elucidate the effects of the multiplicative noise on the retrieval properties. 

\begin{figure}
\begin{center}
\includegraphics[scale=1.0]{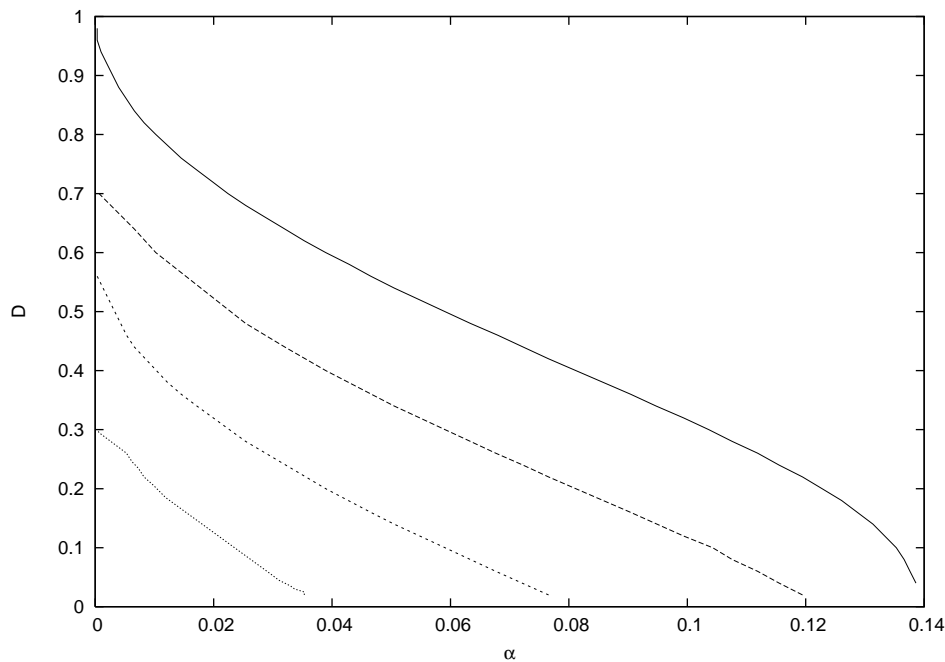}
\end{center}
\caption{\label{fig1}Storage capacity $\alpha$ as a function of the intensity of the external noise $D$ for various values of $\tilde{D}$. 
The solid curve denotes the storage capacity for $\tilde{D}=0$. 
The broken (\broken), dashed (\dashed) and dotted (\dotted) curves represent the storage capacity for $\tilde{D}=0.2$, $0.4$, $0.6$, respectively. 
The retrieval state locates below the curve and vanishes at $\tilde{D}\sim 1.05$. We set $A=20$. }
\end{figure}
\begin{figure}
\begin{center}
\includegraphics[scale=1.0]{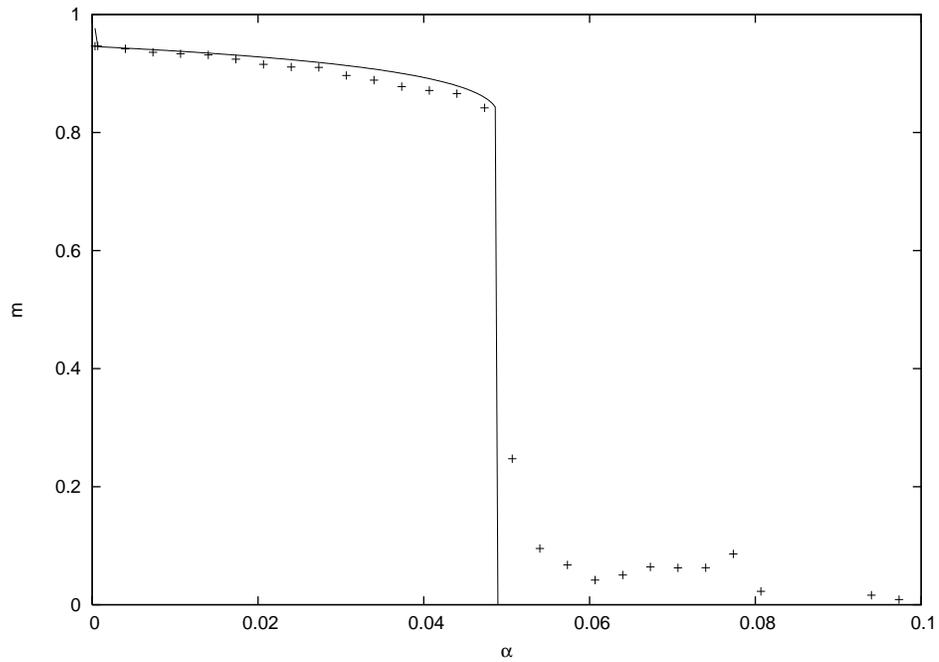}
\end{center}
\caption{\label{fig2}$\alpha$-dependence of the overlap $m^{1}$ obtained from the SCSNA (solid curve) together with that from numerical simulations with $N=3000$ (dots). 
The potential is given by equation (\ref{pot}) with $A=20$.
The intensities of additive and multiplicative synaptic noise are $D=0$ and $\tilde{D}=0.5$ respectively. }
\end{figure}
\begin{figure}
\begin{center}
\includegraphics[scale=1.0]{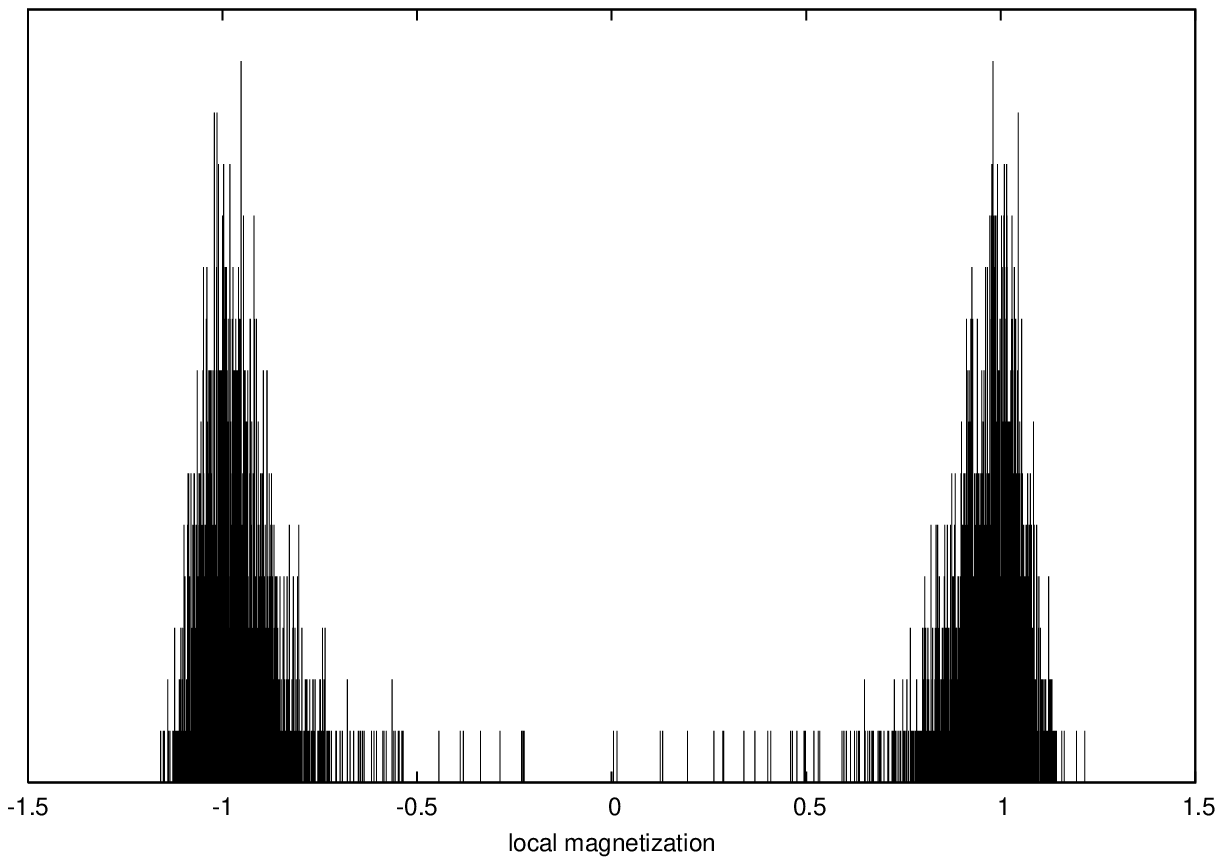}
\end{center}
\caption{\label{fig3}Distribution of the thermal average of the state of neurons, or the local magnetization with parameters $D=0$, $\tilde{D}=0.5$, $\alpha=0.1$, $A=20$. 
From figure \ref{fig2} this parameter set locates in the non-retrieval phase.  
This distribution shows that the non-retrieval spin glass phase arises in this regime. }
\end{figure}

Figure \ref{fig1} illustrates the storage capacity $\alpha$ as a function of the intensity of the external noise $D$. 
The solid line is for the absence of the synaptic noise, i.e., $\tilde{D}=0$. The line in figure \ref{fig1} denotes the numerical solution of the order parameter equations for each intensity of the synaptic noise $\tilde{D}=0$, $0.2$, $0.4$, $0.6$ and the retrieval state vanishes at $\tilde{D}\sim 1.05$ for $A=20$. 
The $\alpha$-$D$ line for the {\it analog network} is deformed compared to the Ising networks. 
This is the effect of potential properties and the {\it effective temperature}, or the {\it non-trivial order parameter} $\hat{q}$, while $\hat{q}=1$ for Ising neurons. 

We can see that the storage capacity is incrementally decreased as the intensity of the synaptic noise increases. 
This result is reasonable since the memories are encoded in the synaptic coupling as local minima of the {\it effective} free energy corresponding to equation (\ref{eqprob}) and the synaptic noise is expected to disturb the fine structure of the energy landscape. 

Figure \ref{fig2} displays the $\alpha$-dependence of the overlap $m^{1}$ obtained from the SCSNA together with that from numerical simulations with $N=3000$. 
We can see that the overlap $m^{1}$ decreases as the number of embedded patterns $\alpha$ increases and retrieval state vanishes ($m^{1}=0$) at $\alpha_{\mathrm{c}}\sim 0.049$. 

Figure \ref{fig3} illustrates the distribution of the thermal average of the state  of neurons, or the local magnetization $\langle x_{i}\rangle$ at $D=0$, $\tilde{D}=0.5$, $\alpha = 0.1$, $A=20$ obtained from numerical simulations with $N=3000$. 
We can see from figure \ref{fig2} that the overlap $m^{1}=0$ in this regime. 
However the local magnetizations $\langle x_{i}\rangle$'s are seen to distribute around $\langle x_{i}\rangle =\pm 1$. 
This means that a spin glass phase arises in this regime. 
We can also show the existence of the non-retrieval spin glass phase analytically by solving the set of order parameter equations (\ref{5.5}), (\ref{sigma}), (\ref{m}), (\ref{U}), (\ref{gamma}) and (\ref{qhat}). 
Since $m^{1}=0$ in the non-retrieval phase, the order parameter equations (\ref{sigma}) and (\ref{U}) become 
\numparts
\begin{eqnarray}
\sigma_{z}^{2}&=&\frac{\alpha}{(1-U)^{2}}\langle F^{2}(z)\rangle_{z}\, ,\label{sigmaSG}\\
U&=&\langle F^{\prime}(z)\rangle_{z}\, ,\label{USG}
\end{eqnarray}
where $\langle\cdot\rangle_{z}$ denotes the average with respect to the Gaussian random variable $z$. 
For the non-retrieval phase $m^{1}=0$, it is trivial that the order parameter equation $m^{1}=0=\langle\xi F(z)\rangle_{\xi ,z}$ holds. 
By definition of the transfer function $F$, the order parameter $U$ is rewritten as 
\begin{eqnarray}
U=\beta_{\mathrm{eff}}(\hat{q}-q)\, ,\label{qSG}
\end{eqnarray}
\endnumparts
where $q=\frac{1}{N}\sum_{i}\langle x_{i}\rangle^{2}$ is the Edward-Anderson order parameter. 
The set of order parameter equations (\ref{5.5}), (\ref{gamma}), (\ref{qhat}), (\ref{sigmaSG}), (\ref{USG}), (\ref{qSG}) takes a closed form. 
Thus we can find the non-retrieval spin glass phase by solving these equations to obtain $q\neq 0$. 
Since the Edward-Anderson order parameter $q$ is expected to be small in the regime close to the paramagnetic phase, the Taylor expansion with respect to $q$ is applicable for these order parameter equations to illustrate the paramagnetic ($q=0$)-spin glass ($q\neq 0$) phase boundary. 
This boundary is expected to correspond to the de Almeida-Thouless (AT) line. 
The study on the relationship between the SCSNA and the replica symmetry breaking is underway. 

\section{Concluding remarks}
We have derived the TAP equation for a stochastic analog neural network with temporally fluctuating multiplicative synaptic noise, which is not found in literatures. 
More specifically, we have derived the TAP equation together with the set of order parameter equations by using the SCSNA and the cavity method. 
Our original model {\it does not} have the concept of free energy. 
Since the self-averaging property holds in the thermodynamic limit $N\to\infty$, we have found that the nonlinear Fokker-Planck equation (\ref{FPE}) becomes quasi-linear to allow one to obtain equilibrium probability density obeying the Gibbs one with effective temperature and hence that the network with white synaptic noise has the {\it effective} Hamiltonian in the large $N$ limit. 
Thus the cavity method, which is applicable to the model with energy concept, becomes available to obtain the (pre-)TAP equation. 
Unlike the case without synaptic noise, the concrete form of the transfer function $F$ of our model has been found to depend not only on the coefficient of the Onsager reaction term but on the order parameter $\hat{q}$. 
$\hat{q}$ as well as the coefficient of the Onsager reaction term have been obtained self-consistently within the framework of the SCSNA. 
The full TAP equation straightforwardly follows from the pre-TAP equation by substituting the solutions of the order parameter equations into the pre-TAP equation (\ref{preTAP}). 

Furthermore, we have found that the storage capacity of the network gradually decreases as the intensity of the synaptic noise increases, since the fine structure of the energy landscape tends to disappear by the interference of the synaptic noise. 
This effect of the interference of the synaptic noise on the behavior of the retrieval property has been shown to appear via the effective temperature $\beta_{\mathrm{eff}}^{-1}\ge D$. 

All the results presented in this paper are obtained via the cavity method and the SCSNA. 
On the other hand, the order parameter equations (\ref{sigma}), (\ref{m}), (\ref{U}), (\ref{gamma}), (\ref{qhat}) can be reproduced as replica symmetric case by the replica method, since the system has the effective Hamiltonian (\ref{hamiltonian}). 
Thus our results are expected to be exact within the replica symmetric approximation. 
However, the development of the analysis in the framework of the SCSNA for replica symmetry breaking solutions is now underway. 

In other works dealing with the temporal fluctuation in synaptic couplings \cite{Marro, Garrido2}, the authors study the case of the fast synapse dynamics. 
Then the synaptic coupling is modified to take the form of "effective synaptic coupling" and the system becomes to have an effective Hamiltonian. 
In this case the "effective synaptic coupling" is {\it straightforwardly determined} by both the number of embedded patterns and the intensity of Langevin noise associated with neuron dynamics. 
On the other hand, the time scale of fluctuation of synaptic coupling in our model is {\it comparable to that of the neuron dynamics}. 
Our model results in having the "effective temperature" and hence the effective Hamiltonian in the thermodynamic limit. 
However, in our model, the "effective temperature" is {\it determined
only self-consistently} together with the other order parameters. 

In this paper we have dealt with a network subjected to asymmetric multiplicative synaptic noise given as white noise involving both pre- and post-neuron and the noise has no correlation with the synaptic coupling given by the Hebb learning rule. 
However some other versions of synaptic noise may be considered: (i) synaptic noise depending only on pre- or post-neuron, (ii) synaptic noise correlated with the Hebb learning rule, (iii) colored synaptic noise. 
For some of these cases, one can rigorously derive the TAP equation and the set of order parameter equations similarly to the case we have seen in this paper. 
The analysis for such cases will be reported elsewhere. 

\ack
This work was supported by a 21st Century COE Program at Tokyo Tech "Nanometer-Scale Quantum Physics" by the Ministry of Education, Culture, Sports, Science and Technology.

\end{document}